Check for updates

*CORRESPONDENCE
Janaka Ekanayake
✉ ekanayakej@cardiff.ac.uk

# The influence of social interactions in mitigating psychological distress during the COVID–19 pandemic: a study in Sri Lanka

Isuru Thilakasiri[1], Tharaka Fonseka[1], Isuri Mapa[2], Roshan Godaliyadda[1], Vijitha Herath[1], Ramila Thowfeek[3], Anuruddhika Rathnayake[4], Parakrama Ekanayake[1] and Janaka Ekanayake[5]*

[1]Department of Electrical and Electronic Engineering, Faculty of Engineering, University of Peradeniya, Peradeniya, Sri Lanka, [2]Faculty of Science, University of Peradeniya, Peradeniya, Sri Lanka, [3]Department of Psychology, Faculty of Arts, University of Peradeniya, Peradeniya, Sri Lanka, [4]Ministry of Health, Colombo, Sri Lanka, [5]Cardiff University, Cardiff, United Kingdom

Massive changes in many aspects related to social groups of different socioeconomic backgrounds were caused by the COVID-19 pandemic and as a result, the overall state of mental health was severely affected globally. This study examined how the pandemic affected Sri Lankan citizens representing a range of socioeconomic backgrounds in terms of their mental health. The data used in this research was gathered from 3,020 households using a nationwide face-to-face survey, from which a processed dataset of 921 responses was considered for the final analysis. Four distinct factors were identified by factor analysis (FA) that was conducted and subsequently, the population was clustered using unsupervised clustering to determine which population subgroups were affected similarly. Two such subgroups were identified where the respective relationships to the retrieved principal factors and their demographics were thoroughly examined and interpreted. This resulted in the identification of contrasting perspectives between the two groups toward the maintenance and the state of social relationships during the pandemic, which revealed that one group was more "socially connected" in nature resulting in their mental state being comparatively better in coping with the pandemic. The other group was seen to be more "socially reserved" showing an opposite reaction toward social connections while their mental well-being declined showing symptoms such as loneliness, and emptiness in response to the pandemic. The study examined the role of social media, and it was observed that social media was perceived as a substitute for the lack of social connections or primarily used as a coping mechanism in response to the challenges of the pandemic and results show that maintaining social connections physically or via online rather than the use of social media has helped one group over the other in decreasing their symptoms such as emptiness, loneliness and fear of death.

## 1 Introduction

The global pandemic, COVID-19 has reported a total of over 776 million cases (COVID-19 Cases, 2024) and over 7 million deaths worldwide (COVID-19 Deaths, 2024) as of April 2024. Since its first identification in December 2019, the virus outbreak affected almost all sectors of society, leaving lasting impacts on a global scale. Throughout the pandemic period, the global population faced challenges in accepting and adjusting to the changes that inevitably affected many aspects of daily life. The "new normal" as it was called, was a phase in which individuals all over the world were adapting to the novel experience in which a majority of the social interactions were limited to a virtual environment owing to the steps taken by governments to mitigate the spread of the pandemic (Kontoangelos et al., 2020). The psychological impact of COVID-19 extended to various other aspects of daily life, owing to the drastic changes that occurred with the travel constraints imposed worldwide (Pedrosa et al., 2020a) and these effects on psychological well-being had lasting impressions even after the infection rates subsided (Saladino et al., 2020; De Sousa et al., 2021).

While social distancing measures have been effective in controlling the spread of COVID-19, they have also diminished people's connection with their social networks (Carpio-Arias et al., 2021; Pancani et al., 2021; Carpio-Arias et al., 2022; Long et al., 2022). In this context, individuals were more likely to increase contact with family members to sustain levels of social support and connectedness which improved psychological well-being during the pandemic (Cooper et al., 2021; Nitschke et al., 2021; Widnall et al., 2022). The depression and anxiety rates have gone up among the global population during the pandemic according to the World Health Organization (2024). In comparison with 2017 reports, it had gone up to levels of 31.5 and 31.9% from 4.4 and 3.6%, respectively, for depression & anxiety during the COVID-19 pandemic (Wu et al., 2021) and the pandemic has left many countries with lasting mental health concerns among the population (Barton et al., 2023; Jamshaid et al., 2023).

Distance learning became the mode of education during the COVID-19 period where this affected students emotionally who were well accustomed to learning with physical guidance (Gurajena et al., 2021; El Refae et al., 2021). Since school students and educators were confined to online learning platforms, increased levels of distress were reported (Liang S. W. et al., 2020; Padrón et al., 2021) and university students had also shown psychological distress (Ghanayem et al., 2024) due to distance learning and its related control measures (Al-Tammemi et al., 2020). Studies showed the necessity to monitor and promote mental health in university students to boost resilience in crisis (Copeland et al., 2021; Padrón et al., 2021). Even though the youth in general was less vulnerable to the COVID-19 disease, they were more at risk considering its negative psychological effects (Liang L. et al., 2020; Orgilés et al., 2020; Power et al., 2020).

The COVID-19 pandemic has had the most negative psychological effects on people who considered themselves as most vulnerable, driven by changes in circumstances such as home confinement, exposure to the disease and disruptions to daily routines (Rodríguez-Rey et al., 2020). Studies aimed at investigating the relationship between the socioeconomic status and mental health outcomes of the pandemic (Agberotimi et al., 2020; Ferrucci et al., 2020) and the general psychological impact (Pedrosa et al., 2020) have been carried out in context specific manners in order to further establish the ramifications COVID-19 has had on this impact area. The rapid spread of the virus on a global scale which resulted in it being one of the major topics of discussion, pandemic related information spread through all forms of media and the drastic measures taken by countries worldwide regarding containment strategies have made the analysis of the pyschological impact of COVID-19 essential. These factors such as fear of the virus, extensive media exposure, and restrictive containment measures have been shown to largely contribute in affecting the psychological well-being of individuals in the global context (Gray et al., 2020; Rodríguez-Rey et al., 2020).

Another major aspect of life that affected human psychology during the pandemic period was the impact on the income and economic level of people (Kikuchi et al., 2021). Regardless of the sector of employment, employees, employers, as well as entrepreneurs, faced challenges such as the risk of unemployment, bankruptcy, and losing customer bases in addition to switching over to completely new platforms and workloads (Ingusci et al., 2021; Phungsoonthorn and Charoensukmongkol, 2022). Individuals who relied on traveling for work often and associates of the tourism industry were severely affected by the travel bans and restrictions causing additional emotional distress (Chen, 2021; Kang et al., 2021). The feelings of job insecurity, and being unable to provide for families on top of the changes that occurred in the economy worldwide created a sense of unpredictability and uncertainty. This contributed to the psychological impact of the pandemic on individuals increasing anxiety, especially in low-income countries such as Sri Lanka (Alonzo et al., 2021; Kola et al., 2021; Senarath et al., 2024).

The switch to virtual connectedness resulted in drastic changes being made to the professional careers of individuals as well (Kniffin et al., 2021). In addition to the heightened feelings such as anxiety related to their job insecurity, people needed to adapt to the novel methods recommended in delivering their services (Kontoangelos et al., 2020). The limitations in technical literacy added to the stress associated with work specifically in older generations that needed to completely depend on the usage of virtual platforms and online resources in their respective fields of employment (Chen, 2021; Maduwage et al., 2021). Even during partial lockdown periods where social distancing was practiced, the fear of infection the virus caused uneasiness and anxiety in individuals who had to leave households (Lin et al., 2021). Essential workers and healthcare professionals were a group that was severely affected psychologically, owing to their constant exposure to the pandemic and high-risk associated work (Giusti et al., 2020; Momenimovahed et al., 2021; Sun et al., 2021).

During the COVID-19 pandemic, restrictions and quarantines led to a massive increase in the use of forms of entertainment and social media globally (Zhao and Zhou, 2021; Aggarwal et al., 2022). Social media emerged as a powerful tool during this time, introducing many people to its use for the first time (González-Padilla and Tortolero-Blanco, 2020). Those previously unfamiliar with social media began to adopt it, while those with some prior exposure, especially the youth, significantly increased their usage (González-Padilla and Tortolero-Blanco, 2020). With the shift to distance learning, young people had constant access to devices for their classes and lectures, providing an easy gateway to social media (Drouin et al., 2020). People turned to social media as a coping mechanism (Kotozaki et al., 2023; Social Media and Social Anxiety, 2022; Orsolini et al., 2022), however, its use had both positive and negative consequences (Venegas-Vera et al., 2020; Lelisho et al., 2023). Some studies suggested that social media use helped mitigate the feeling of loneliness during the pandemic (Marciano et al., 2022) where on the other hand social media also served as a medium that spread false news, panic and uncertainty (Abbas et al., 2021). Social connectedness and social media were a topic of discussion (Pandya and Lodha, 2021) as mentioned above, even if social media use was associated with factors that negatively impacted psychological well-being, studies have shown that negative impacts could be lessened if proper measures are taken (Ostic et al., 2021). Furthermore, through our study, it was intended to evaluate whether social media acted as a mechanism to alleviate the lack of social connectedness during the pandemic or whether it was more aligned with the characteristics of the methods that people use as coping mechanisms such as substance use and food.

Considering the effect of the pandemic on South Asian countries other than Sri Lanka, a study showed that two-fifths of the Indian population were experiencing anxiety and depression due to lockdown (Grover et al., 2020) while many people welcomed this pandemic with a belief that this would make doctors and researchers well-prepared for future pandemics and shape politicians into more responsible (Bhattacharya et al., 2021). The associated uncertainty has been increasingly testing the psychological resilience of not only the general public but also healthcare professionals (Raj et al., 2020; Sandesh et al., 2020). Similar to the global scenario, in South Asia, social connectedness and family relationships played a vital role in helping the general public cope with the challenges of the pandemic (Pandey et al., 2021; Rehman et al., 2021).

In the South Asian region, the overall psychological impact was adverse and the prevalence of depression, anxiety, sleep disorders, and alcohol use disorders increased (Mia and Griffiths, 2022). About one-third of the students who participated in a study in the South Asian region experienced anxiety during the pandemic (Chinna et al., 2021). Students had adopted coping strategies such as religious practices to tackle the psychological issues (Salman et al., 2022; Mishra et al., 2023). University students experienced high levels of psychological distress while transitioning to the new norms due to COVID-19 (Dhahri et al., 2020; Islam et al., 2020; Faisal et al., 2022) and during home quarantine, students experienced excessive stress from both their academic workload and the isolation, putting their mental health at serious risk (Khan et al., 2020).

Since the identification of the first case of COVID-19 in Sri Lanka in March 2020, the government implemented stringent travel restrictions, nationwide lockdowns, and social distancing measures to effectively block the virus's spread throughout the country (Epidemiology Unit, 2024). There have been studies focused on the psychological state of different specific groups in Sri Lanka. Fear of being infected with COVID-19 or spreading it among family members were major risk factors that seem to cause depression and anxiety for healthcare professionals during the pandemic (Perera et al., 2021; Udayanga et al., 2022) and parents faced stress during the lockdown due to resource-poor situations (Athapathu et al., 2022). During this period of stress and uncertainty, social media was poorly managed in Sri Lanka compared to other countries and this situation led to the spread of false news and fear, causing unnecessary panic among the population (Akuratiya and Akuratiya, 2021; Jayarathne and Wijesinghe, 2023). Studies indicate that people's lifestyles were significantly influenced by their social media usage during this period (Jinasena and N'Weerasinghe, 2022). University students faced major challenges such as concerns with their grades, challenges with adapting to the virtual education systems, workload and time management in an online learning environment, and as a result, a stressful environment for distance learning was created affecting their mental health (Madhusanka et al., 2021; Lakmini and Jayathunga, 2023). Furthermore, a significant increase in depression and anxiety was present among students with a history of psychiatric disorders (Rohanachandra et al., 2021; Baminiwatta et al., 2022). A study conducted across five faculties in the University of Ruhuna showed a significant level of stress, anxiety and depressive symptoms among undergraduates (Kaushani and Weeratunga, 2023). The unavailability of resources and issues in connectivity in certain areas, specifically in low-income countries such as Sri Lanka, deprived students of receiving a consistent education during the period of self-isolation which often resulted in their mental well-being being negatively affected and the average income of the household had been identified as the key denominator of separation of the students in Sri Lanka (Senarath et al., 2024). These studies were done focusing on very specific social groups and it was evident that the fear of infecting the virus, misleading news in social media, online mode of education delivery, and uncertainty could be identified as the common factors which affected the mental state of these specific social groups.

Sri Lanka can be considered as a suitable candidate to study the impact of the pandemic psychologically, as a result of its diverse population and ability to analyze data from multiple groups of individuals belonging to various socio-economic backgrounds. Furthermore, a lack of data-driven studies aiming to analyze the impact of the pandemic psychologically is present in most low-income countries including Sri Lanka. In addition to the lack of such studies done in such low-income countries, there is a research gap in using unsupervised methods to identify social groups that were similarly affected during the pandemic. The use of unsupervised methods can conclude without prior judgment and that can be considered of great importance in identifying the characteristics of human behavior. A comprehensive summary of the contributions of this study is elaborated below.

The data utilized in this study was obtained through a nationwide door-to-door, face-to-face, CAPI (Computer-Assisted Personal Interview) field survey, which includes the necessary level of in-depth interactions essential for an extensive impact analysis. Face-to-face field surveys are more effective for data collection due to their ability to include responses from vulnerable communities, those with limited resources or computer literacy, those with privacy concerns and especially for psychological impact assessments (Hillier et al., 2014). Therefore, our study has employed a more efficient method to record the first-hand, authentic experiences of Sri Lankans which ensures a comparatively better analysis of the psychological impact of the pandemic especially considering the level of interaction needed in capturing emotions and personal experiences accurately. This study was done not based on a prior hypothesis but based on a data-driven approach to identify how the mental state of the population was affected and the dataset used in this study is digitalized and published online (Ilangarathna et al., 2023, 2024). As such, the data can also be used in future work pertaining to psychological state assessment studies using data engineering and artificial intelligence (AI). Research studies in psychology that incorporate AI are still relatively limited.

However, interest in integrating machine learning and deep learning into these studies has grown over time and our research also employs AI techniques to draw conclusions from the data (Yarkoni and Westfall, 2017; Dwyer et al., 2018). These techniques offer significant potential for effective analysis in the social sciences as well.

This study employed principal component analysis (PCA) based factor analysis and unsupervised clustering to identify key underlying factors which affected the psychological state of the Sri Lankan population using the above-mentioned dataset. After the identification of the common factors which affected the general population, the data points corresponding to each household were categorized into distinct groups using unsupervised clustering which is a popular approach used in machine learning, enabling us to draw insightful conclusions. Although unsupervised clustering methods are relatively uncommon in traditional statistical analysis, they are highly popular in machine learning-based AI systems, reflecting the broader AI revolution (Lindsay, 2020). The conclusions drawn from these techniques are further explained in the results section.

We believe that the findings of this research have the potential to provide valuable insights and information to evaluate the psychological impact of COVID-19 on different socio-economic groups, allowing government agencies, policymakers, counseling practitioners to make informed decisions, construct solutions to the psychological issues that might arise in a future pandemic. In addition, it will also benefit mental health professionals in terms of helping them understand how different population groups responded to the stresses associated with the pandemic, particularly their distinctive nature in dealing with the associated psychological impacts. Understanding these differences will aid in deciding on more effective, proactive treatment strategies. This study also provides insights for future extended research within this field inviting further validation by the broader academic and policy-making communities.

## 2 Materials and methods

### 2.1 Dataset and participants

The data collection process of this research consisted of a nationwide door-to-door, face-to-face, CAPI (Computer-Assisted Personal Interview) field survey (Ilangarathna et al., 2023) conducted covering 3,020 households in Sri Lanka from November 6, 2021, to December 10, 2021. With a population of approximately 22 million and an estimated 6.76 million households (Preliminary Report, 2024), the required sample size for this study was determined using a 95% confidence interval and a 2% margin of error. Since the population is large and the variability is unknown, maximum variability was assumed by setting $p = 0.5$. Using the formula for sample size calculation for proportions, the required sample size was calculated to be 2,401 (Cochran, 1977; Hu and Bentler, 1999). To enhance representativeness and ensure robustness, the survey covered a total of 3,020 households. These households were selected from all nine provinces and they represented 20 out of the total of 25 districts in the country. Households were selected using a multistage cluster sampling technique where factors such as population, severity of the risk of disease spread were taken into consideration and the selection was done reflecting the hierarchical power distribution of administrative divisions in Sri Lanka. The sampling framework was formed using data from the Department of Census and Statistics of Sri Lanka and was further supported by village-level administrative officers. Consequently, the selected households were representative of the population's diversity, encompassing various ethnicities, age groups, employment sectors, and other demographic characteristics. Table 1 compares the Socio-demographic characteristics of the total survey dataset and the processed dataset used for this study. Figure 1 illustrates a comparison between the percentage distributions of the surveyed data and Sri Lankan population data based on district population statistics.

Furthermore, a thorough review of the population statistics, economic standing, and high-risk locations based on the region's dependency ratio was conducted in the process of selecting the households in a region and to ensure that the data well representative, areas with urban, rural, agricultural, industrial, fisheries, and estate sectors were also included. The household selection procedure is further explained in Figure 2. The corresponding Village Officer assisted in selecting the households that were most severely impacted by the pandemic, such as households that faced more distress compared to the neighborhood, families that dealt with extremely poor financial situations, and households where the entire family got infected.

The survey questionnaire which took around 40 min delving into the impact of the pandemic on these households featured a comprehensive questionnaire with approximately 78 questions, along with 8 additional questions on basic demographic information. Responses were collected using a Likert scale with numerical values assigned to each unique response. This face-to-face field survey approach proves to be the more successful method of data collection in comparison to an online survey due to its ability to include responses from individuals belonging to more vulnerable communities in society who may experience a lack of resources or computer literacy required to participate in an online survey (Hillier et al., 2014; Zhang et al., 2017) and even other individuals who may opt out of responding to an online questionnaire owing to privacy concerns and unfamiliarity with the entities conducting the survey (Hlatshwako et al., 2021; Wardropper et al., 2021). The questions related to the basic demographic information were answered by the main respondent of each household while questions related to each impact area were answered by its most relevant respondent. Gender, relationship to the head of the household, age, marital status, education level, and ethnicity, were collected as the basic demographic information (Ilangarathna et al., 2024). The individual areas studied by the survey are mobility and human behavior, income and economic status, food consumption, education, access to health services & related information, and cultural and psychological changes. The survey addressed impact areas with the following number of questions: 22 questions focused on educational impact, 14 on access to health services and related information, 12 on income and economic status, the impact on food consumption was discussed in 6 questions, and 7 on mobility and human behavior. Additionally, cultural impact was covered by 5 questions and psychological changes by 12 questions.

The responses for the 12 questions in the psychological impact section (Ilangarathna et al., 2023) (section 4D of the survey) of the questionnaire were recorded with the use of a Likert Scale, varying from 1 to 6, where these, respectively, translated into "No change, Slightly increased, Markedly increased, Slightly decreased, Markedly decreased, Cannot say." This method of defining the scale was selected

TABLE 1 Socio-demographic characteristics of the study population total survey dataset and the processed dataset used for this study.

| Variable | Survey population | | Processed dataset | |
|---|---|---|---|---|
| | *n* | % | *n* | % |
| | 3,020 | 100 | 921 | 100 |
| **Gender** | | | | |
| Female | 885 | 29.30 | 107 | 11.62 |
| Male | 2,135 | 70.70 | 814 | 88.38 |
| **Respondents' relationship to the head of the household** | | | | |
| Head of the household | 2,420 | 80.13 | 850 | 92.29 |
| Wife/Husband | 507 | 16.79 | 51 | 5.54 |
| Son/Daughter | 65 | 2.15 | 18 | 1.95 |
| Parent | 18 | 0.60 | 1 | 0.11 |
| Other relatives | 10 | 0.33 | 1 | 0.11 |
| Non relatives | 0 | 0 | 0 | 0 |
| Other (Specify) | 0 | 0 | 0 | 0 |
| **Age** | | | | |
| 20–30 years | 221 | 7.32 | 62 | 6.73 |
| 31–40 years | 574 | 19.01 | 190 | 20.63 |
| 41–50 years | 851 | 28.18 | 273 | 29.64 |
| 51–60 years | 708 | 23.44 | 207 | 22.48 |
| 61–70 years | 465 | 15.40 | 142 | 15.42 |
| 71–80 years | 181 | 5.99 | 47 | 5.10 |
| 81–90 years | 20 | 0.66 | 0 | 0 |
| **Marital status** | | | | |
| Unmarried | 80 | 2.65 | 26 | 2.82 |
| Married | 2,655 | 87.91 | 853 | 92.62 |
| Widowed | 250 | 8.28 | 39 | 4.23 |
| Divorced | 17 | 0.56 | 2 | 0.22 |
| Separated | 18 | 0.60 | 1 | 0.11 |
| **Education level** | | | | |
| None | 133 | 4.40 | 24 | 2.61 |
| Pre school | 3 | 0.10 | 0 | 0 |
| Grade 1 - Grade 5 | 270 | 8.94 | 67 | 7.24 |
| Grade 6 - G.C.E O/L | 1906 | 63.11 | 549 | 59.61 |
| G.C.E. A/L | 560 | 18.54 | 215 | 23.34 |
| Diploma | 41 | 1.36 | 15 | 1.63 |
| Degree | 100 | 3.31 | 45 | 4.89 |
| Masters or higher | 7 | 0.23 | 6 | 0.65 |
| **Ethnicity** | | | | |
| Sinhalese | 1796 | 59.47 | 673 | 73.07 |
| Tamil | 592 | 19.60 | 102 | 11.07 |
| Muslim | 623 | 20.63 | 144 | 15.64 |
| Mixed | 2 | 0.07 | 1 | 0.11 |
| Other | 7 | 0.23 | 1 | 0.11 |

*(Continued)*

TABLE 1 (Continued)

| Variable | Survey population | | Processed dataset | |
|---|---|---|---|---|
| | *n* | % | *n* | % |
| **Employment status** | | | | |
| Government employee | 257 | 8.51 | 138 | 14.98 |
| Semi government employee | 39 | 1.29 | 15 | 1.63 |
| Private sector employee | 630 | 20.87 | 320 | 34.74 |
| Estate worker | 62 | 2.05 | 23 | 2.50 |
| House wife/unpaid family worker | 140 | 2.64 | 8 | 0.87 |
| Unemployed | 633 | 20.96 | 0 | 0 |
| Self (specify) | 160 | 5.30 | 47 | 5.10 |
| Retired | 155 | 5.13 | 77 | 8.36 |
| Daily wage earner | 200 | 6.62 | 84 | 9.12 |
| Other (specify) | 744 | 24.64 | 209 | 22.69 |
| **Employment sector** | | | | |
| No relevant response | 792 | 26.23 | 0 | 0 |
| Health sector | 46 | 1.52 | 24 | 2.61 |
| Education sector | 95 | 3.15 | 44 | 4.78 |
| Retail sector | 376 | 12.45 | 137 | 14.88 |
| Manufacturing sector | 211 | 6.99 | 82 | 8.90 |
| Bank sector | 21 | 0.70 | 8 | 0.87 |
| Forces | 143 | 4.74 | 79 | 8.58 |
| Administrative services sector | 99 | 3.28 | 36 | 3.91 |
| Apparel sector | 81 | 2.68 | 37 | 4.02 |
| Agriculture sector | 161 | 5.33 | 85 | 9.23 |
| Other (specify) | 995 | 32.95 | 389 | 42.24 |

since the responses relevant to the psychological impact area were targeted to identify the emotions, behavior and relationships of the respondent. However, for the analysis, this Likert scale was relabeled and converted into a symmetrical one to achieve an accurate representation from the analysis and the changed values of the initial scale and the scale used in the analysis are given in Table 2. The responses with the answer "Cannot say" were processed in two different ways resulting in two different analyses. Firstly, the responses with the answer "Cannot say" were dropped (as shown in Table 2) and in the second analysis, the responses with the answer "Cannot say" were relabelled as zero (0) denoting those answers are also considered as "No Change" answers. Both these methods yielded very similar results and considering the most accurate representation being the first method, i.e., dropping the "Cannot say" responses in the process of converting the scale into a symmetrical one, and considering the results were similar for both approaches, results of only the first approach are presented in this study.

The initial step of the exploratory data analysis was to select the relevant responses from the original dataset, which resulted in the processed dataset consisting of 921 responses out of the total of 3,020 (30.5%) responses. This selection of only the relevant responses can be summarized in three steps as follows. Firstly, the households that gave no responses to at least one of the questions given in the psychological impact section of the questionnaire (0 values in the dataset) were removed resulting in 2651 entries out of 3,020 (87.78%), followed by the second step of removal of entries that had errors in collection or entering processes (denoted by 99) which then resulted in a dataset with 1969 responses (65.2%). Thereafter, the responses with "Cannot say" as an answer were dropped, resulting in 921 responses (30.5%) for the final dataset ($n = 921$). As mentioned above, the survey was designed to collect information on multiple avenues such as mobility and human behavior, economic status, food consumption, education, health services, and not just the psychological impact. Therefore, after the data cleaning, we can be certain that all of these 921 households had complete and accurate information related to psychological impact analysis. The questions in the survey pertaining to psychological impact were selected based on the results of a literature review (Ilangarathna et al., 2023, 2024) and an initial premise that the psychological state would have been impacted by factors such as the harmony of the household, financial situation, work-life balance, increased exposure to social media, lack of peer interaction, loneliness. In order to include as many questions as possible to cover the totality of the psychological effect analysis, these questions were chosen using a vast spanning criterion.

As shown in Figure 1 a higher proportion of survey respondents were found in certain districts, which also happen to be the most densely populated districts in Sri Lanka compared to the overall Sri Lankan population. This was primarily because more groups that were severely affected by the pandemic were present in these districts than in the rest of the nation. The purpose of the survey was to ensure that the data distribution was as representative as possible, while gathering the most amount of information possible from the groups that were mostly affected by the pandemic. The Kaiser-Meyer-Olkin (KMO) reliability measure was tested against the data used in this study and this test assesses the suitability of data for factor analysis by measuring sampling adequacy for individual variables and the entire model (Watson and Thompson, 2006; Wetzel, 2012; Hou, 2023) and the values are given in the Results section.

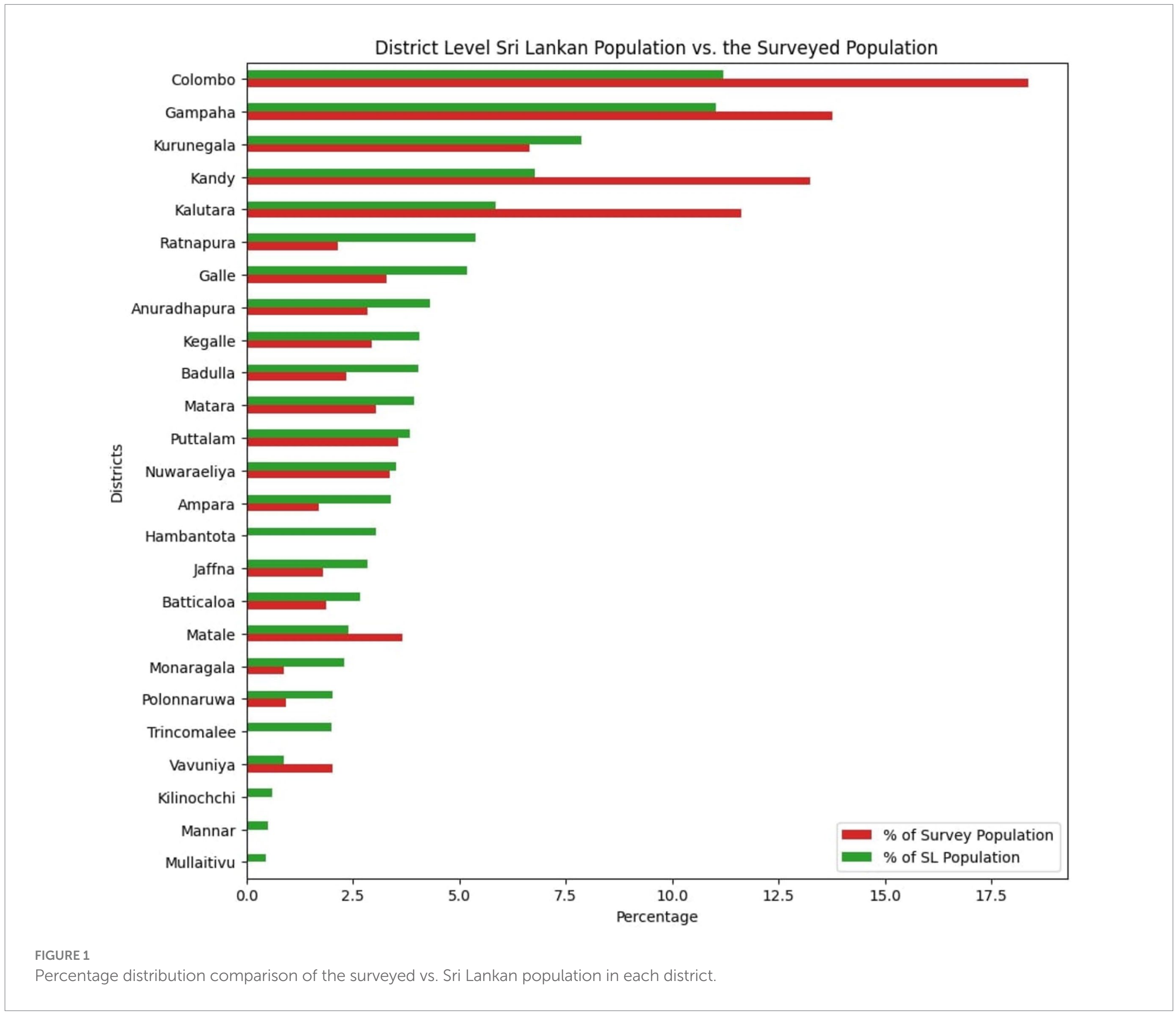

FIGURE 1
Percentage distribution comparison of the surveyed vs. Sri Lankan population in each district.

## 2.2 Factor extraction and number of factors

In order to identify the latent factors in a multivariate dataset, factor extraction methods play a crucial role. In this study, SPSS 27 statistical analysis software was utilized to extract the factors from the dataset with PCA and Varimax rotation. Factor analysis reduces the dimension space into a set of vectors that condenses the effect of the individual questions. These reduced directions encompass several questions depending on their contribution to each reduced dimension. The number of dimensions in which the dataset must be reduced is selected using certain criteria and this study considered certain thresholds for this selection process according to the literature. One widely used approach is the Kaiser Criterion, which relies on eigenvalues to identify the number of factors. In PCA, larger eigenvalues indicate higher variance explained by the data. According to this criterion, factors are identified as any variable with an eigenvalue greater than 1 is considered as a factor and this was considered in this study (Watson and Thompson, 2006; Wetzel, 2012; Acar Güvendir and Özer Özkan, 2022). Subsequently, Varimax rotation was applied to the factor loadings obtained through PCA by considering a threshold for the cutoff factor loading value as 0.4 (Watson and Thompson, 2006; Bavel et al., 2020; Hou, 2023). The cutoff value of 0.4 is selected to minimize the cross-loading effect in the rotated component matrix. Cross-loading items represent the questions that have a factor loading more than 0.4 under more than one factor with loading differences less than 0.1 (Muilenburg and Berge, 2005). This occurs when a question is significantly contributing to the direction of two factors in the considered space even if the underlying idea requires them to be significantly contributing to only one direction. Therefore, such

**Provinces**
9 provinces in Sri Lanka. Survey was conducted encompassing all nine provinces.

**Districts**
20 out of Sri Lanka's 25 districts were selected for the survey, taking into account the severity of the COVID-19 spread and the containment strategies implemented during the pandemic.

**Divisional Secretariate (DS) Divisions**
89 out of 331 Divisional Secretariat (DS) divisions was selected for the survey. The chosen DS divisions represented a diverse cross-section of Sri Lankan society, including urban, rural, estate, industrial, fisheries, and agricultural areas, to minimize bias in the data.

**Grama Niladhari (GN) divisions**
14,022 Grama Niladhari (GN) divisions in Sri Lanka, a sample of 215 was chosen for the survey. These GN divisions were selected based on high-risk criteria, as determined by a dependency ratio map produced by the Department of Census and Statistics. The map identifies areas with a high proportion of elderly and young populations. Additionally, GN divisions with significant fisheries, agriculture, estate, and industrial sectors were prioritized.

**Households**
A total of 3,020 households, each with an average of four members, were selected for the survey. These households were identified as severely impacted by the pandemic, experiencing multiple hardships such as widespread infection or severe financial strain. Local administrative officers assisted in identifying such households, and a random component was introduced to the selection process to ensure diverse representation of the affected population.

FIGURE 2
Household selection procedure of the survey.

questions are removed from the factor analysis to achieve the most accurate factor analysis and to simplify the factor loading arrangement. This removal of questions can mitigate any bias that could be present inherently in the questionnaire itself, created during the question formulation. Once such questions are removed, the factor analysis with PCA and the rotated component matrix calculation is repeated until the aforementioned assumptions of threshold for factor loadings and threshold for cross-loadings are met.

After identifying the number of factors, a new space was created using these factors, where each point in this space represented a respondent. This space was then analyzed to determine if there were any groups that were similarly affected. The clustering process involved two steps: first, spectral clustering was employed to determine the optimal number of clusters, and then this number was used as input for the k-means algorithm to cluster the population.

TABLE 2 Change in scale performed during data preprocessing.

| Description | Initial scale [used in the survey] | Converted scale before processing the data |
|---|---|---|
| No change | 1 | 0 |
| Slightly increased | 2 | 1 |
| Markedly increased | 3 | 2 |
| Slightly decreased | 4 | −1 |
| Markedly decreased | 5 | −2 |
| Cannot say | 6 | Dropped from the study (Explained above) |

## 2.3 Spectral clustering

Spectral clustering is a machine learning technique frequently employed when data-driven approaches are warranted, particularly for its ability to identify clusters in data that may not be linearly separable. According to literature, spectral clustering has also been a successful technique for determining clusters in complex datasets, demonstrating robust performance in various applications where traditional clustering methods may fall short (Chen et al., 2017; Guo et al., 2020; Senarath et al., 2024). The projected dataset resulting after the dimension reduction step of the factor analysis can be clustered to identify the data points or the individuals representing those data points who have been affected similarly. Therefore, the big single cluster lying in the high-dimensional space which was created by combining several sub-clusters, can be clustered using several different techniques. When zooming in on the cluster space, sub-clusters start to form within the feature space and this mechanism of zooming in to detect the number of small clusters submerged inside the initial supercluster is called the "Modes" of clustering (Qin et al., 2016; Rupasinghe et al., 2016; Van Dam et al., 2021). Additionally, the number of clusters in a mode acts as the unique identifier for that mode. Therefore, different modes are identified by adjusting the free parameter "$\sigma$," which represents the zooming effect in the Standard Spectral Clustering algorithm. This procedure is referred to as "Sigma Sweep." The primary steps of Sigma Sweep are as follows:

1 Affinity matrix a, for the refined dataset was generated using the transformation

$$A_{i,j} = \begin{cases} \exp\left(-\dfrac{\|x_i - x_j\|^2}{2\sigma^2}\right) & ;\ i \neq j \\ 0 & ;\ i = j \end{cases}$$

Where $\sigma$ is a tuneable parameter,

2 Degree matrix D is a diagonal matrix where $(i,i)^{th}$ element of D is the row sum of A s' $i^{th}$ row.

$$D_{i,k} = \begin{cases} \sum_j A[i,j] & i = k \\ 0, & \text{Otherwise.} \end{cases}$$

3 Calculate the laplacian matrix L, from the relationship

$$L = I - D^{-1/2} A \ D^{-1/2}$$

Where $I$ is an identity matrix.

4 Obtain eigenvalues of L and arrange them into descending order. Then eigengaps were computed by getting the difference between two successive eigenvalues.
5 Plot the graph of the variation of the eigengap with log σ by repeating steps 1–4 for different σ values to determine the number of sub-clusters K.

To determine the ideal number of clusters, K, sigma sweep characteristics are utilized. The figure demonstrating log σ vs. eigengap from spectral clustering, as shown in Figure 3, indicates that the second eigengap predominates over a wider range of sigma values, suggesting the presence of two clusters in the space.

Once the number of clusters is determined using the spectral clustering algorithm, this cluster count can be used as the input for the 'k' value in the K-means algorithm. Subsequently, K-means can be applied to cluster the population.

## 2.4 K-means clustering

K-means clustering algorithm is a widely used unsupervised algorithm for partitioning a dataset into a pre-defined number of clusters (K) (Krishna and Narasimha Murty, 1999; Ikotun et al., 2023).

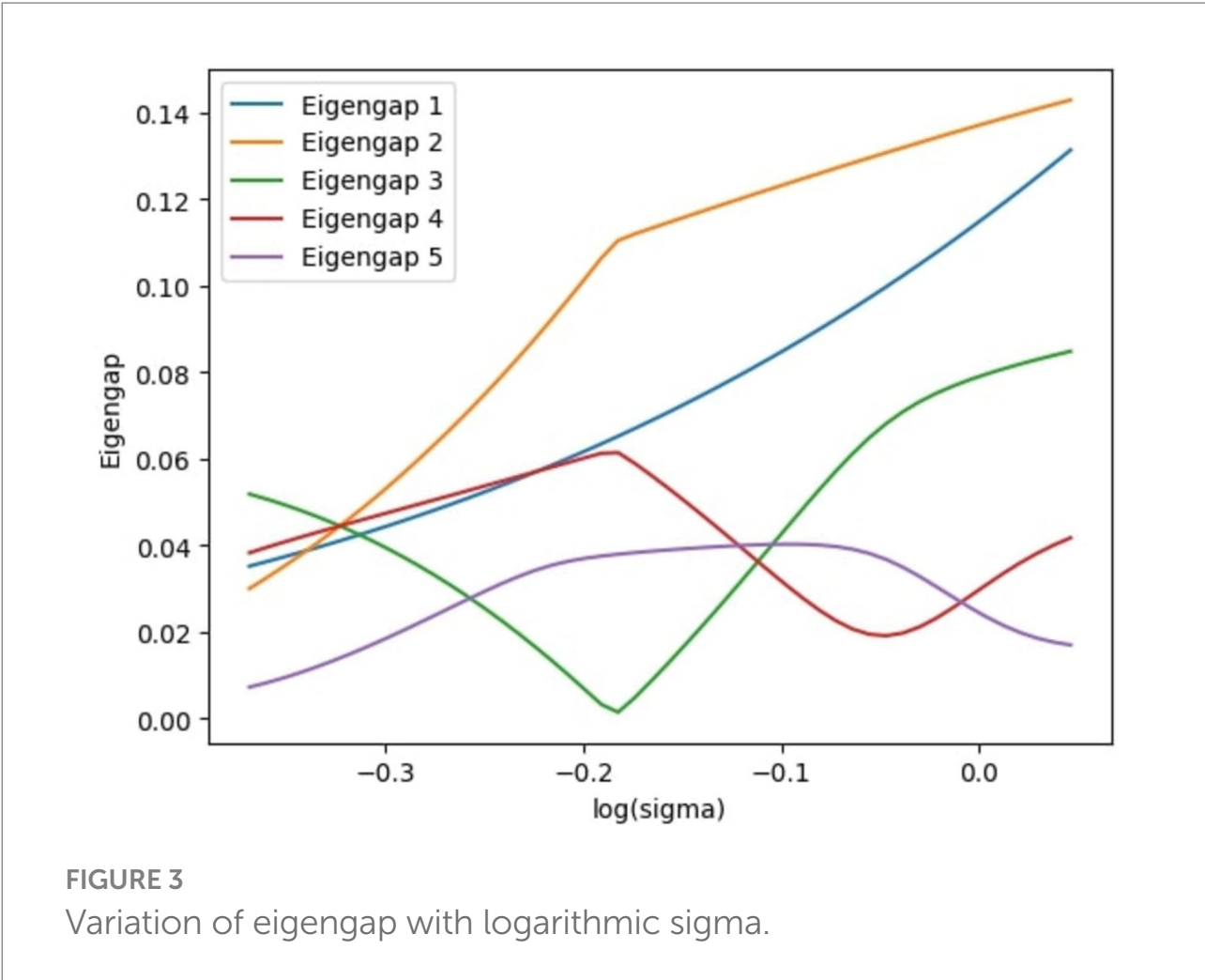

FIGURE 3
Variation of eigengap with logarithmic sigma.

The aim is to cluster similar data points together and observe the underlying patterns. This is carried out by minimizing the distance between the points within a cluster. K-means algorithm is identified as a centroid-based algorithm where each cluster is associated with a centroid. If the goal is to find K clusters, K centroids must be defined and that can be done manually or randomly depending on the computer program used. Through an optimization process, the best set of centroids that minimizes the sum of squared distances between each data point and its closest centroid will be found. Considering the random method, initially, K random points in the dataset will be selected and they act as the initial cluster centroids. Then, for each data point in the dataset, the distance between that point to each K centroid will be calculated and the data point will be assigned to the cluster whose centroid is closest to it. This is where it effectively forms K clusters. Once all points have been assigned to clusters, recalculation of the centroid is carried out by taking the mean of all data points assigned to each cluster. Thereafter, the same process will be continued for the new centroids until convergence. The convergence occurs when the centroids no longer change significantly or when a specified number of iterations are reached. In this study, K-means clustering was applied to the reduced space of the dataset with to four dimensions after PCA with Varimax rotation. Using such a step, it was intended to cluster the survey population using an unsupervised method into groups of people who were affected by the pandemic similarly.

# 3 Results

This study presents the outcomes of three data-driven components of analysis. Firstly, the factors identified using factor analysis that predominantly contributed toward the psychological impact of the COVID-19 pandemic on the population of Sri Lanka. These factors provided interesting insights into how the population has coped and perceived different forms of social contact, and social media. Secondly, the population was clustered using unsupervised clustering to identify the number of groups and the participants that fall under these identified groups to recognize the participants who were similarly psychologically affected by the pandemic. Specifically, during the lockdown periods, relationships with other individuals were affected severely and it was observed that the perception of different relationships has evolved during the pandemic period in Sri Lanka. Finally, we have discussed how the Sri Lankan population was affected during the pandemic by evaluating the demographics of the people that fall under the clusters, analyzing their relationship with the underlying factors recognized by the factor analysis and considering how the people in each cluster have answered the survey.

As mentioned above in the Materials and Methods section, the questions corresponding to each identified factor were selected in accordance with the threshold introduced in existing literature where only questions with a factor loading above 0.4 were considered as eligible to be representative of each factor. After checking the cross-loading effects with a 0.1 threshold, and rectifying them, the process was repeated until such effects were no longer present.

The factor analysis of this study resulted in only one iteration where only one question had to be removed from the analysis in accordance with the thresholds mentioned in Materials and Methods section. The "Sleep" question was removed in the above-mentioned iteration and that resulted in the final analysis data with 11 variables (11 questions) and the mean value answers for each of these 11 questions are shown in Figure 4. Only factors with eigenvalues greater than 1 were considered during the factor analysis resulting in four distinct factors as explained further in the Materials and Methods section. The eigenvalues of the four factors were, respectively, 3.0785, 1.5372, 1.2956, 1.0936 and the eigenvalue corresponding to the 5th factor was 0.8207. The cumulative explainable variance of the dataset with four factors was 63.681% of the total variance. Figure 5 shows how the cumulative explainable variance of the dataset varied across the 11 components and the scree plot for the factor analysis is illustrated in Figure 6. Rotated component matrix consisting of the questions corresponding to each factor from the factor analysis with their respective factor loadings are shown in Table 3. The nature of the factors was identified after considering the question groups under each factor and these factors were identified as "Coping Mechanisms, Symptoms, Work-Life Balance, Peer Interactions / Connections." An overview of all the 12 questions and where they belonged after the factor analysis is shown in Table 4. The Kaiser-Meyer-Olkin Measure (KMO) values were 0.749 and 0.703, respectively, for the dataset with all questions considered and for the dataset after dropping the "Sleep" Question.

Furthermore, complementary factor analysis (CFA) was conducted to confirm and validate the factor structure. The fit indices obtained for our model indicated a good overall fit, with values such as Goodness of Fit Index (0.962), Adjusted Goodness of Fit Index (0.93), Comparative Fit Index (0.925), and Normed Fit Index (0.912) exceeding the recommended threshold of 0.9, reflecting the model's ability to accurately represent the data structure. The low Standard Root Mean Square Residual (0.033) and acceptable Root Mean Square Error of Approximation (0.072) further confirm that the residuals are minimal and the model effectively captures the relationships among variables. These high values suggest that the hypothesized relationships align well with the observed data, supported by an adequate sample size, high data quality, and appropriate estimation methods. However, the results from CFA indicated a Chi-square over Degrees of Freedom value of 5.72 ($\chi^2$/df = 5.72), which exceeds the commonly recommended threshold of 5 for an acceptable model fit (Rex, 1998). This may primarily be due to the large sample size, which is known to inflate chi-square values and make it overly sensitive to minor discrepancies (Rex, 1998). Based on these indices and their acceptable ranges, the model can be considered a good fit (Hu and Bentler, 1999).

Figure 4, which denotes the mean values of the responses given in the processed dataset, stands as an attestment to the universal experience the population of Sri Lanka collectively went through during the lockdown period. As observed in the above-mentioned figure, the "Relationship with family members" has increased in the overall population of respondents which is understandable considering that most households served as quarantine places for families during the months of imposed lockdowns and travel restrictions. In contrast, it can be observed that the relationships with office colleagues, neighbors and social connectedness have decreased in the overall respondent population as the mean answers have shown negative values in Figure 4 for the "Relationship with Office," "Relationship with Neighbors," "Social Connectedness" questions. Maintaining effective interpersonal relationships during an era of social distancing, quarantine and virtual meetings was certainly a challenge and the decrease shown in the mean values of the dataset related to these questions justifies the experiences of most individuals in terms of upholding interactions with the society and peers extending beyond the immediate family residing alongside them. It must be also noted that "Social Connectedness" does not necessarily

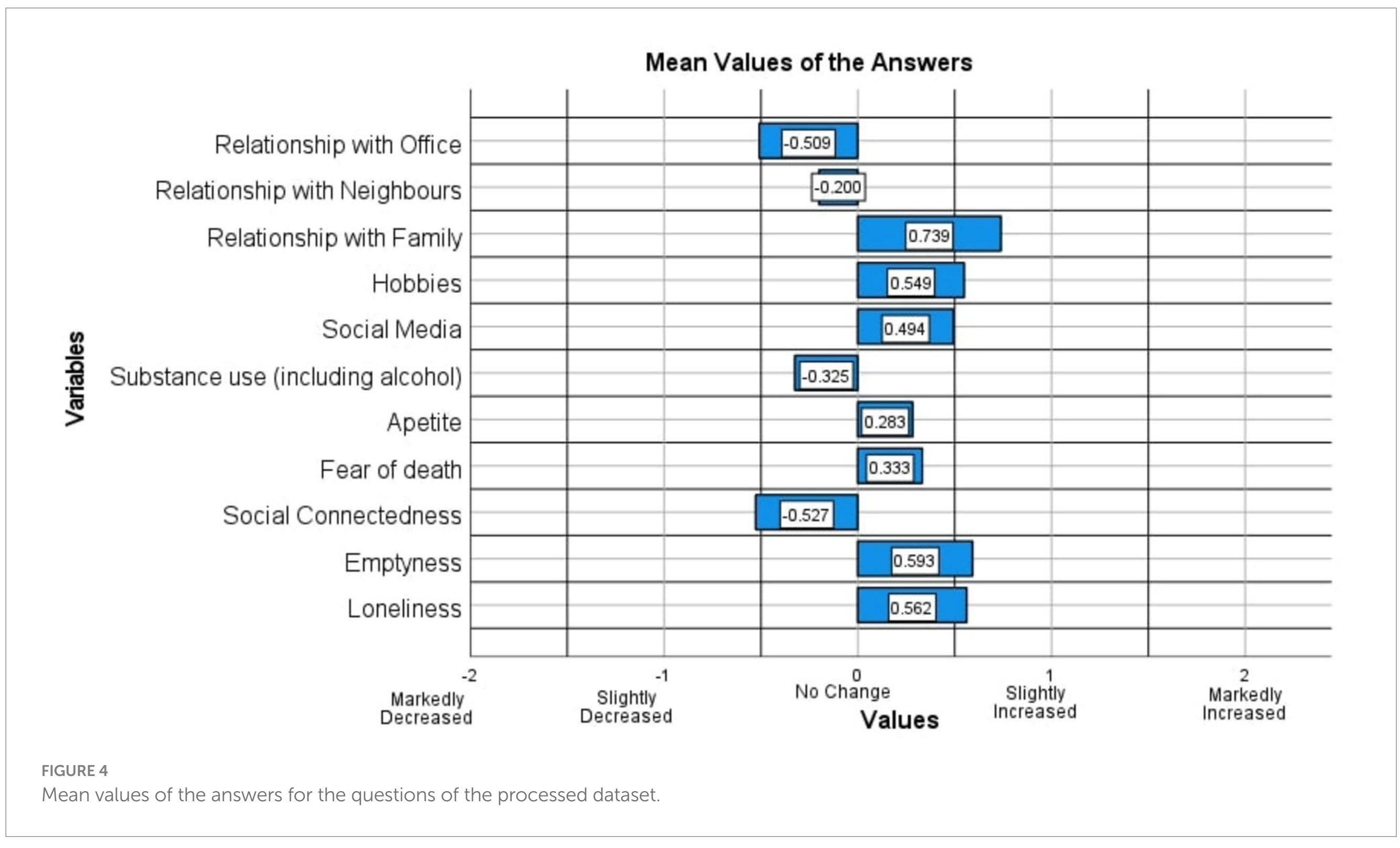

FIGURE 4
Mean values of the answers for the questions of the processed dataset.

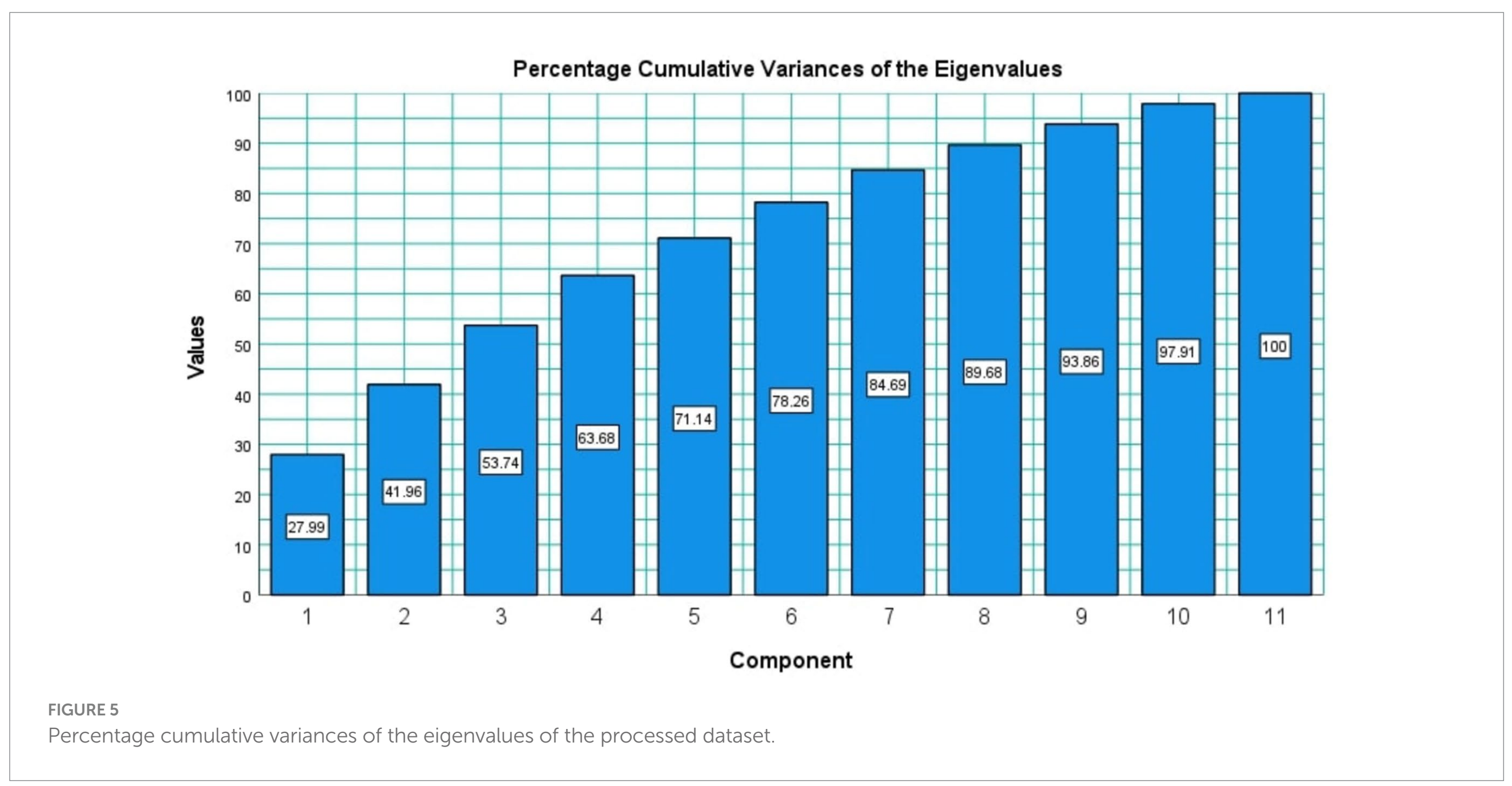

FIGURE 5
Percentage cumulative variances of the eigenvalues of the processed dataset.

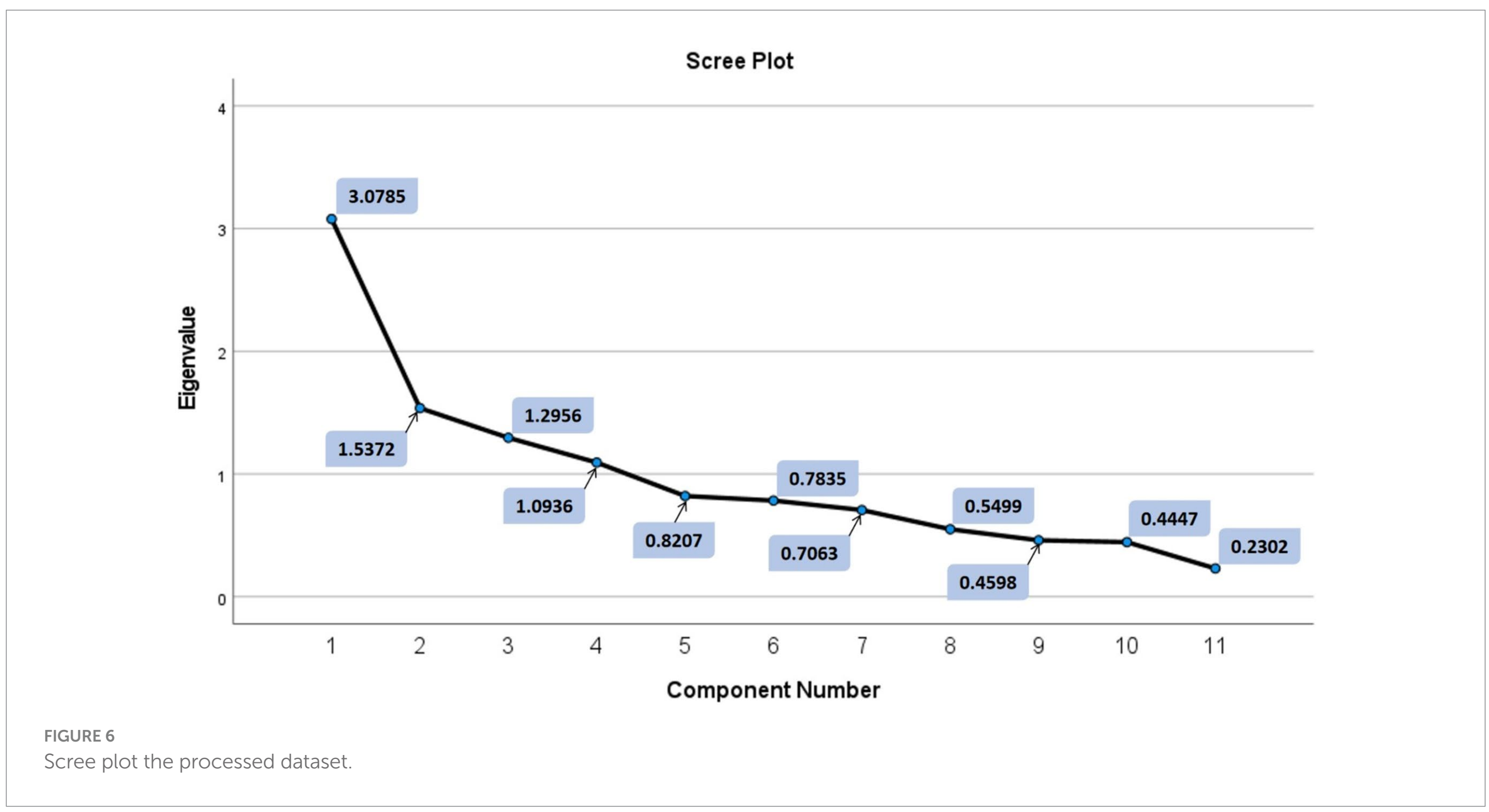

FIGURE 6
Scree plot the processed dataset.

mean only physical interactions, but also voice calls, and video chats which allowed people to have conversations with their close ones.

Furthermore, the symptoms of emotional distress such as feelings of emptiness, loneliness and fear of death have also shown an overall increase when considering the responses for the questions "Emptiness," "Loneliness," and "Fear of Death." This behavior of data is justifiable given the situation that existed during the pandemic period where people were isolated and confined to their homes, forced to a completely novel lifestyle. An increase is visible in the mean of responses to questions related to the engagement in "Hobbies," use of "Social Media" and "Appetite," which is sensible in the same context as mentioned previously where individuals confined to households with monotonous routines tend to seek the lost control, they previously had over their lives using various other mechanisms they can exert an influence on. Therefore, the behavior of the mean values of data corresponding to the responses given in the processed dataset as illustrated by Figure 4, further justifies the nature of psychological experiences and behavior of individuals during the lockdown, quarantine and other stages of the pandemic.

As mentioned above, the factor analysis resulted in the identification of four distinct factors, later named as "Coping Mechanisms," "Symptoms," "Work-life balance" and "Peer interactions/connections." The questions regarding "Hobbies, Appetite, Social Media and Substance Use including alcohol" were grouped under the first identified factor, "Coping Mechanisms" while the questions related to "Loneliness, Emptiness and Fear of death" were categorized under the factor appropriately named "Symptoms." It was also observed that the questions corresponding to the relationship maintained with office colleagues along with the relationship with family members were correlated to each other under "Work-Life Balance." The final factor named "Peer Interactions/ Connections" consisted of the questions discussing the relationship with neighbors and social connectedness. Figure 7 shows the loadings related to "Social Media" question and interestingly, it was highly correlated toward "Coping Mechanisms," rather than "Peer interactions/ connections." Therefore, it can be observed that people perceived social media as a form of coping mechanisms rather than a platform for social connections and this is further elaborated in the Discussion section below.

TABLE 3 Rotated component matrix.

| | Component | | | |
|---|---|---|---|---|
| Variable/Question | Factor 1 | Factor 2 | Factor 3 | Factor 4 |
| Hobbies | 0.735 | | | |
| Appetite | 0.712 | | | |
| Social media | 0.622 | | | |
| Substance use | −0.613 | | | |
| Loneliness | | 0.814 | | |
| Emptiness | | 0.798 | | |
| Fear of death | | 0.672 | | |
| Relationship with office | | | −0.819 | |
| Relationship with family | | | 0.613 | |
| Relationship with neighbors | | | | 0.847 |
| Social connectedness | | | | 0.775 |

TABLE 4 Summary of the questions corresponding to each factor from factor analysis.

| Factor number | Question code | Variable/Question | Factor name |
|---|---|---|---|
| Factor 01 | 4D_1_9 | Watching movies, playing indoor games, reading books/Drawing/Painting | Coping mechanisms |
| | 4D_1_6 | Appetite | |
| | 4D_1_8 | Use of social media | |
| | 4D_1_7 | Substance use including alcohol | |
| Factor 02 | 4D_1_1 | Loneliness | Symptoms |
| | 4D_1_2 | Emptiness | |
| | 4D_1_4 | Fear of death | |
| Factor 03 | 4D_2_3 | Relationship with office colleagues | Work-life balance |
| | 4D_2_1 | Relationship with family members | |
| Factor 04 | 4D_2_2 | Relationship with neighbors | Peer interactions/ Connections |
| | 4D_1_3 | Social connectedness | |
| Dropped questions | 4D_1_5 | Sleep | |

After the factor analysis, the dataset was clustered using unsupervised clustering methods to identify how the pandemic had affected different classes of people and to recognize the key demographic features which the clusters are separated. To identify the number of appropriate clusters, spectral clustering was used where the dominant mode corresponds to the number of clusters. After spectral clustering, it was evident that the maximum eigengap was present between two and three, (as shown in Figure 3), which suggested that ideally there are two clusters. Therefore, people (i.e., the data points) belonging to the two clusters were identified by using the K-means clustering and the K-means clustering diagram is shown in Figure 8.

Subsequently, the average answers the people in the two clusters had given for the questions and the impact of the identified factors on the two clusters was analyzed. This analysis provided valuable insights about the basis on which the clusters had been separated during the unsupervised clustering. From the results shown in Figures 9, 10, it was concluded that the clustering has been done depending on the "socially connected or reserved" nature of the respondent and this has been discussed in length in the Discussion section. Therefore, it was evident that the common denominator for the population in terms of the psychological impact of the pandemic was how well they maintained their social connections. It is justifiable since the connected or reserved nature of an individual is a major factor which determines how well a person can cope with the change in circumstances during the pandemic era with travel and other forms of restrictions. It was identified that people in Group 01 were more "socially connected" and the people in Group 02 were more "socially reserved." Their behavior is further explained in the Discussion section.

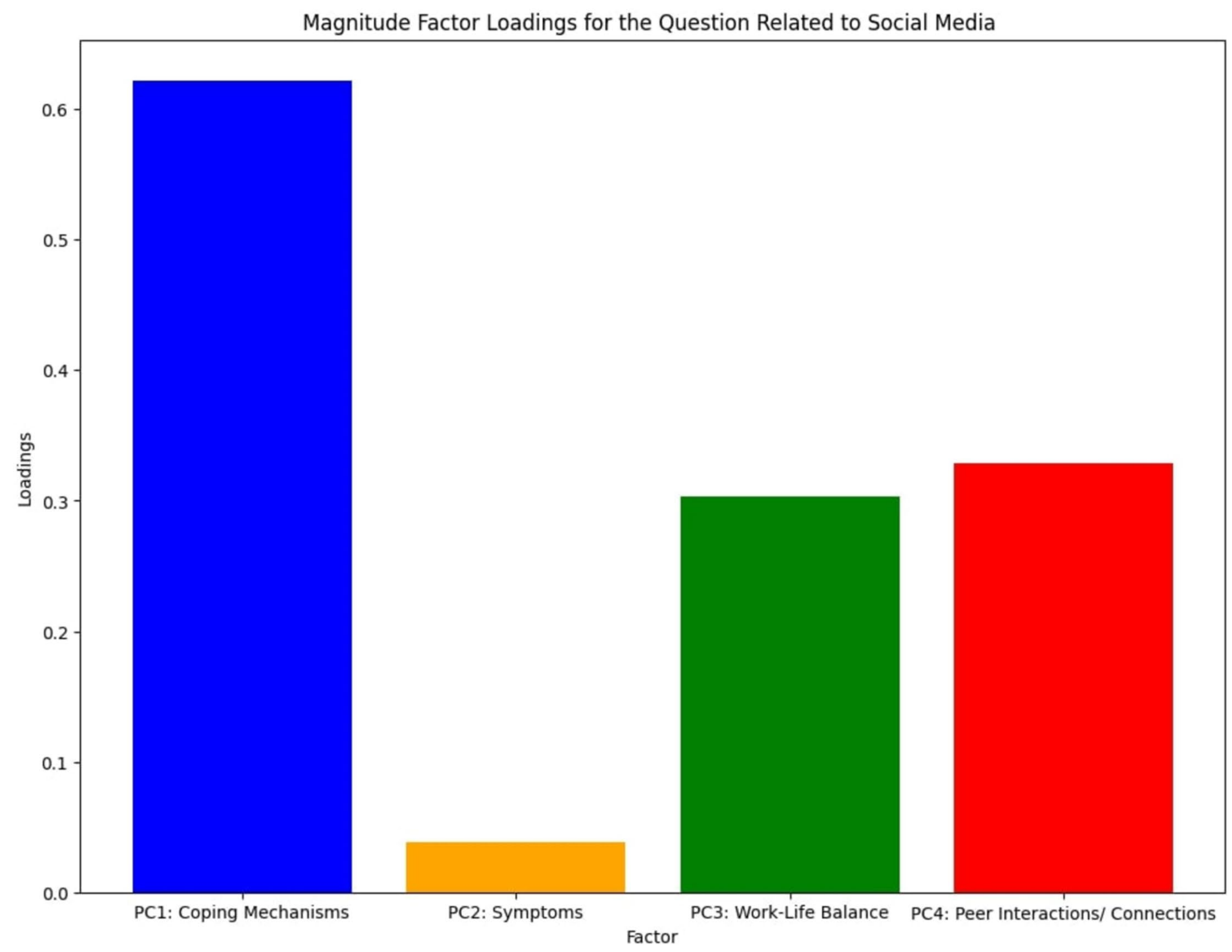

FIGURE 7
Magnitude factor loadings for the question related to social media.

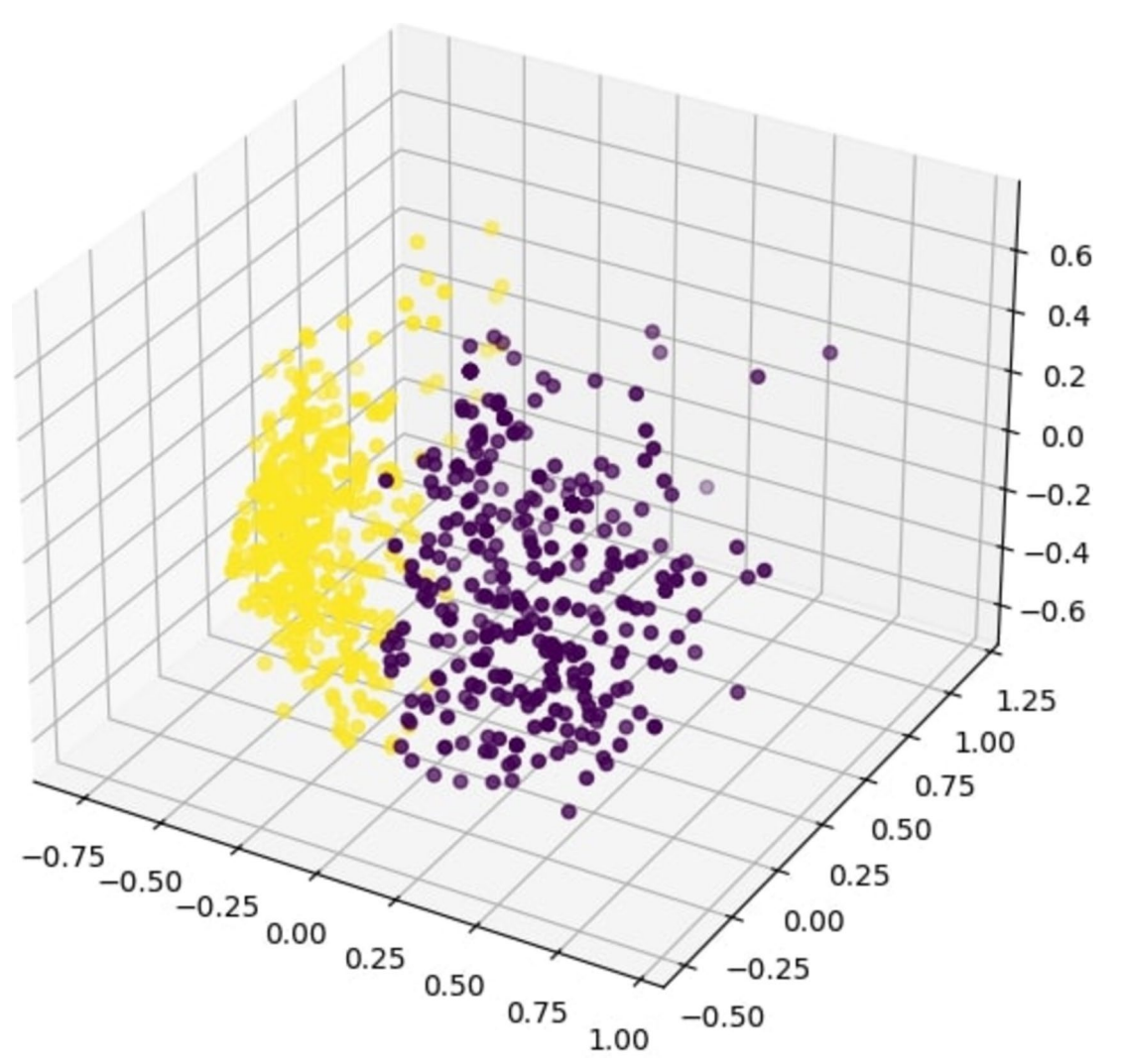

FIGURE 8
Separation of the clusters resulted from unsupervised clustering.

The demographic information collected from the survey (Ilangarathna et al., 2024), was used to identify other characteristics of the people grouped into the two clusters. Namely, the district which the household is located in, gender of the respondent, respondents' relationship to the head of the household, respondents' age, respondents' marital status, respondents' education level, respondent's ethnicity, and respondents' average income level during the COVID-19 pandemic were used for the aforementioned analysis. However, a clear denominator was not present considering any of the above-mentioned demographic information and the only evident separation was present depending on their connected or reserved nature as mentioned previously.

## 4 Discussion

Identifying the factors which affected the psychological state of the population and identifying the demographical and behavioral differences of the social groups that can be identified from the said analysis were the main objectives of this study. A comprehensive analysis was done in order to effectively understand the underlying reasons behind the grouping of the questions under each identified factor from the factor analysis as shown in Tables 3, 4. The first factor "Coping Mechanisms" was identified to correspond with the engagement of individuals in recreational activities, forms of stress release and forms of entertainment during the lockdown period. When we take a look at which questions are grouped under the first factor, it can be concluded that the population have perceived food, hobbies and substance use including alcohol as coping mechanisms during the unprecedented time of self-isolation and travel restrictions that took place during the COVID-19 pandemic. This is in line with the other studies done during the pandemic where the same was observed (Kotozaki et al., 2023; Eden et al., 2020; Cauberghe et al., 2021).

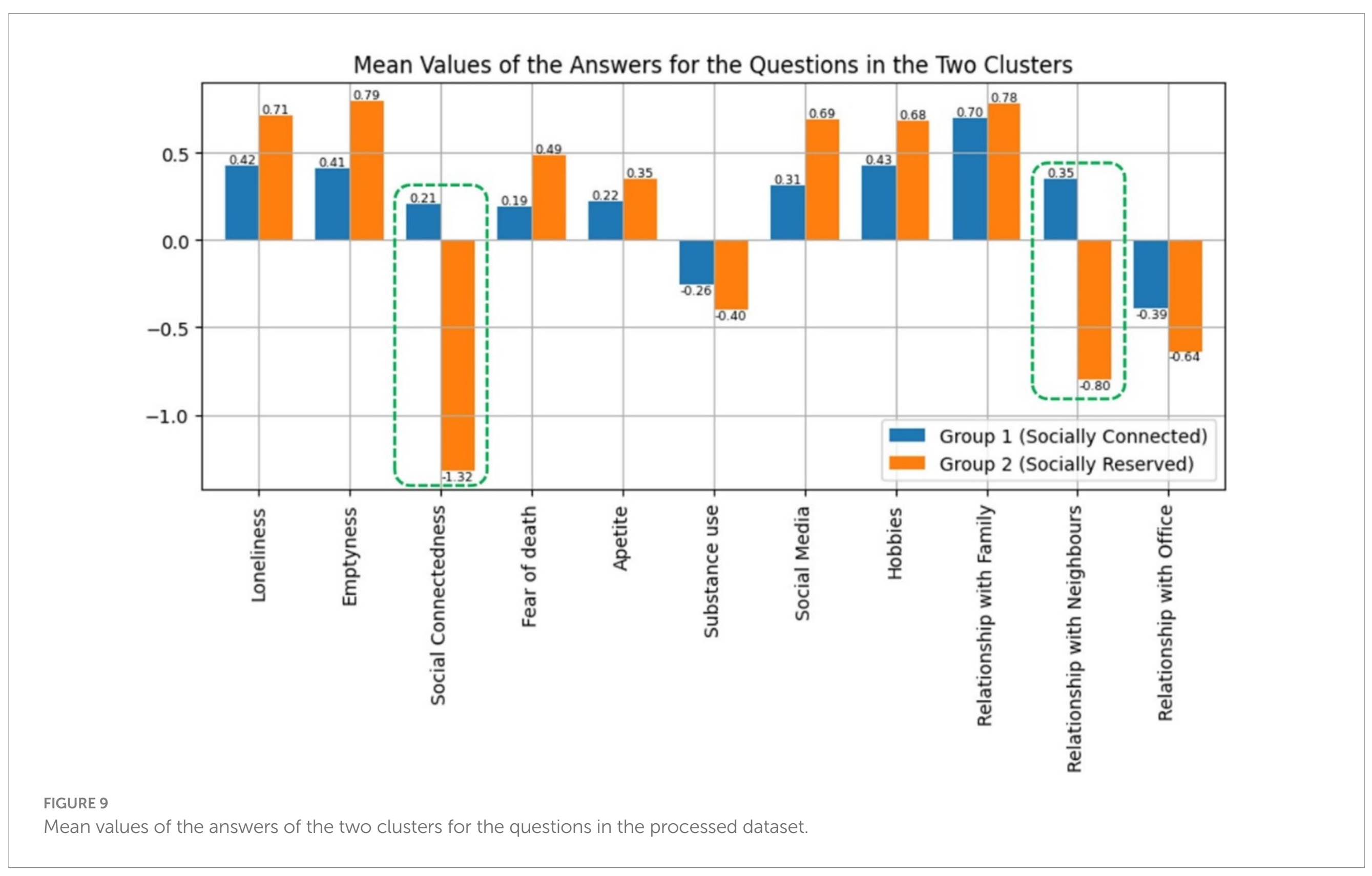

FIGURE 9
Mean values of the answers of the two clusters for the questions in the processed dataset.

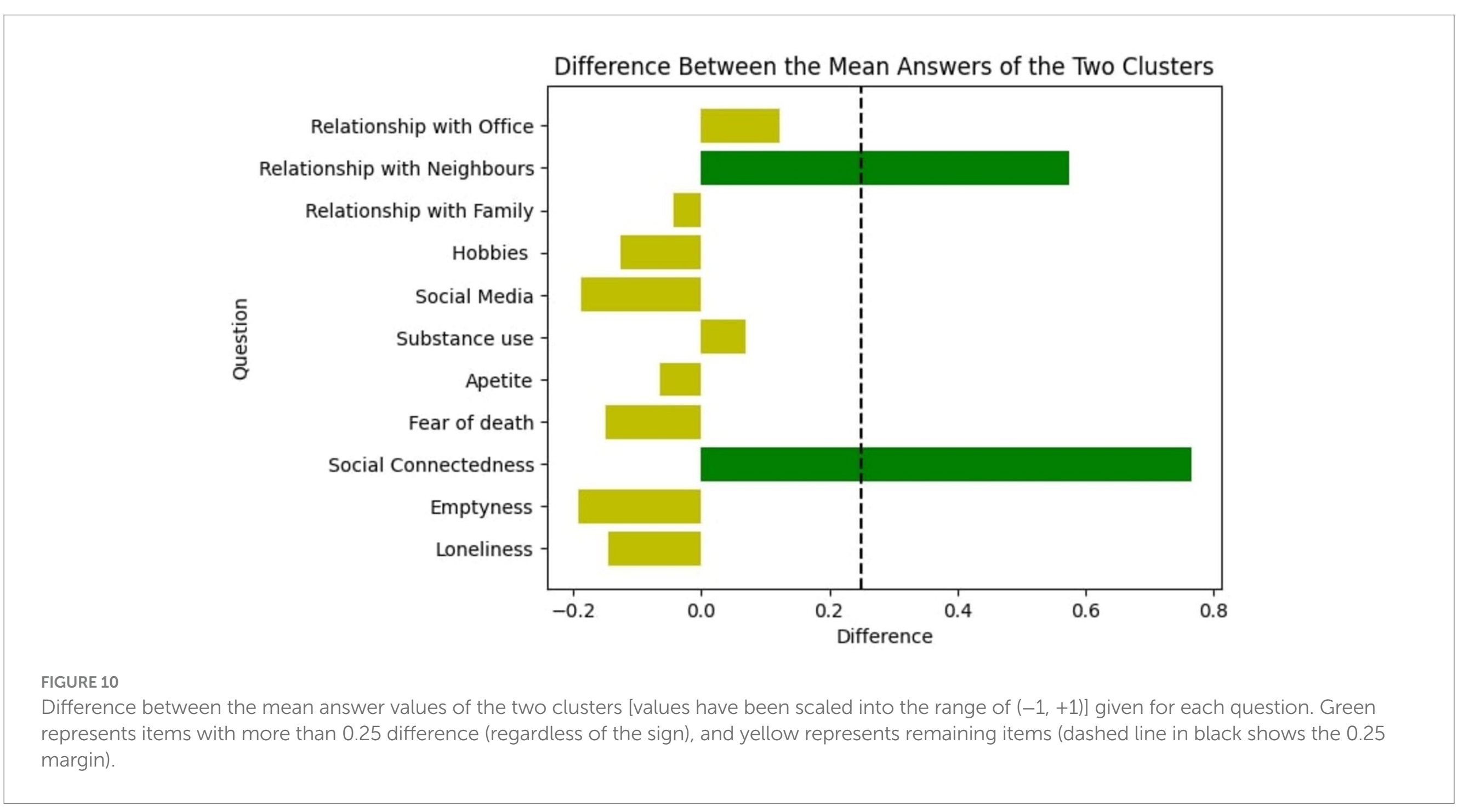

FIGURE 10
Difference between the mean answer values of the two clusters [values have been scaled into the range of (−1, +1)] given for each question. Green represents items with more than 0.25 difference (regardless of the sign), and yellow represents remaining items (dashed line in black shows the 0.25 margin).

One of the most important observations of the factor analysis, specifically considering the first factor, is that most respondents have not perceived social media as a form of peer interaction or as a proper way of continuing human connections. Instead, they have considered social media to be a form of coping mechanism for combating loneliness during self-isolation and also as a source of entertainment since "Social Media" is not grouped with the question related to the state of interpersonal connections (factor loadings shown in Figure 7). The areas or mechanisms related to questions that were grouped under the "Coping Mechanisms" factor can also be identified as the most common behavioral addictive elements of an average person in the 21st century (Miller, 2007; Labib et al., 2022).

This observation of observation of social media acing as a mere distraction, is in line with recent studies done focusing on the impacts of the COVID-19 pandemic (Hajek and König, 2021; Ostic et al., 2021b). During the pandemic, many people could not interact with their peers face-to-face and certain groups of people did not opt to use at least social media for this purpose due to the identified disadvantages of using social media leading them in to an even greater state of mental distress (Ostic et al., 2021). This is consistent with the results of this study where good social interactions made a positive impact on the psychological well-being of the population as further discussed below. However, some studies have shown a connection between increased connectedness and social media among considering the older adult population in the United States (Yu et al., 2016) which shows that when the individuals are less capable of maintaining physical interactions with others due to disabilities or considering their senior age, social media platforms provide feelings of increased connectedness. On the other hand, Generation Z have suffered from social media fatigue and have shown a trend toward discontinuing social media (Liu et al., 2021). However, this attempt to discontinue social media has been reduced by the fear of missing out (Jaspal and Breakwell, 2022), which goes in line with the above observation of perceiving social media as a form of entertainment and a distraction, rather than a tool for maintaining proper human connections (Jaspal and Breakwell, 2022).

The finding of social media not being perceived as a proper form of peer interaction is supported by the factor analysis. Specifically, "Relationship with neighbors" and "Social connectedness" were clustered under factor 04 (Peer interactions / Connections), while "Social Media" was clustered under a different factor (factor 01 – Coping mechanisms). The widespread use and dependency on social media today often reflects mindless browsing and living in a self-controlled virtual reality, rather than being meaningful, emotion-driven interactions with society. Supporting our findings, other research has indicated that social media exacerbates stress and loneliness among university students, rather than providing a platform for social interaction (Ghanayem et al., 2024) where interpersonal relationships are required for better mental health (Bellis et al., 2020). This is further established from the results shown in Figure 10 as the more "socially reserved group" has had increased levels of loneliness, emptiness, and fear of death where they have turned more toward coping mechanisms such as social media rather than trying to maintain social connections.

Considering the third factor from the factor analysis, we can observe that their contributing values have taken opposite signs. This goes on to show that the relationship the general population had with office colleagues and with their family members has acted as opposite elements in the intended measured construct during the pandemic. This can be explained since most workers were asked to carry on their jobs in virtual mode except for the essential services such as health and transportation and keeping a well-executed balance between your work responsibilities and your family responsibilities was paramount. Most of the population had to adapt to the new normal of handling both sides at the same time and at the same place. This has been a key area of focus during the pandemic for a majority of the families therefore, considering the variables, the third factor was named, "Work-Life Balance" (Ingusci et al., 2021; Kniffin et al., 2021; Lonska et al., 2021). The working class was identified as a susceptible group of people and had shown signs of decreased performance in job related activities with decreased amounts of close-proximity interactions with co-workers (La Rosa et al., 2024). How the family cooperated with the members who worked during the pandemic had also impacted the productivity, unity inside the family (La Rosa et al., 2024) and both these findings go in line with the results of this study regarding psychological distress/ social connectedness.

On the other hand, this can also be further explained by relating to three other cases of work-related circumstances. Firstly, the state of the relationship the employed household members had with their workplace was a defining factor since many lost jobs during this time due to the financial downfall of certain sectors (Brewer and Gardiner, 2020). Secondly, even when someone was not at immediate risk of losing their job, the ongoing state of their employee-employer relationship also affected how the rest of the household had to interact with the employed members of the house since the incidents, changes or repercussions in the workplace automatically affected how they behaved in the household as well (Lonska et al., 2021). Thirdly, considering daily-wage workers, many of them faced hardships just to stay afloat, cope with the new circumstances and still find work to earn money (Almeida et al., 2021) and considering all these scenarios, we can justify the above observations related to the third factor.

Questions related to self-isolation, loneliness, emptiness and fear of death were grouped together under the factor 02 and since these are symptoms of psychological distress, this factor was named "Symptoms." This behavior of data could be comprehensively justified given the psychological context of experiences that all groups of society had to undergo, where most individuals were confined to their households having to undergo a period of quarantine willingly or unwillingly (Pancani et al., 2021). This situation was true during the period in which the data was collected as well (Ilangarathna et al., 2024). Although this was a necessary and justifiable action taken by the policymakers and relevant authorities, it is understandable to see an increase in such negative feelings given that humans are naturally wired to behave as social creatures and engage in meaningful interactions with other human beings (Goodwin et al., 2020). On the other hand, during lockdowns with travel restrictions, when individuals were confined to stay at home and work, it is understandable that the feelings of loneliness, and emptiness could occur more often (Ernst et al., 2022). The fear of infection, i.e., perceived health risk among adolescents had negatively impacted them and the symptoms visible through their behavior had a connection to how well they coped with the measures that could be taken such as social media, musical instruments and entertainment (La Rosa and Commodari, 2023). The university students were also in contrast about the future of their respective fields and how well they could cope with the changes in future job prospects, job security and financial stability. Due to these negative psychological changes, the students were more inclined to depend on their online presence and social media (La Rosa and Commodari, 2023) which provides further basis to the findings of this study. Furthermore, even when people had the opportunity to go out under strict conditions, they were at risk of encountering individuals who might have been unknowingly infected or who might have been in close contact with an infected person. Moreover, the fear of death may have existed in the household as a whole, with the risk of exposure not only limited to one member of the household traveling outside but also because they are also residing under the same roof (Gundogan and Arpaci, 2024). Therefore, as the results show, the fear of death has been closely linked as one of the unhealthy psychological symptoms people showed during the pandemic (Pérez-Mengual et al., 2021). This is further established since the average answer shows an increase in these symptoms for the "Loneliness," "Emptiness," "Fear of Death" questions as shown in Figure 4.

In addition to this, another explanation could focus on individuals who closely tied their identity to their professions. During the time period, there existed a heightened sense of job insecurity, especially among most private sector employees, as the prospect of a return to normalcy seemed distant. This feeling was further heightened by the various types of misleading information circulating through news and social media (Apuke and Omar, 2021). Therefore, it can be assumed that this uncertain period of time may have ignited a form of identity crisis among individuals, who may have experienced a confusing state of realizing their existence, self-worth, and future and rediscovering themselves during the period of isolation at home with little to no engagement in their professions. Even in relation to the individuals who worked remotely or essential service-related professionals who continued their work, it is understandable that a sense of doubt might have existed within them owing to the way in which they perceived the change in situation (Kroencke et al., 2020).

Next, considering the two distinct clusters identified following the unsupervised clustering of the processed dataset, revealed interesting results related to the psychological impact of the pandemic that affected the population of Sri Lanka. Figure 9 shows the mean answer values the people corresponding to the two clusters have given to each of the questions. The difference in responses between the two groups in the aforementioned mean values is illustrated in Figure 10, which has been scaled into the range $[-1, +1]$ for better representation. Figure 10 provides better clarity on how the mental state of people in the two clusters changed, how they approached change in circumstances and how they perceived the pandemic time. Considering the comparison of responses further illustrated in Figure 10, we can conclude that the two identified clusters have shown distinguishable characteristics. To elaborate this point, we can identify that the two groups show a considerable difference when it comes to their approach in social connections when we compare the answers for "Social connectedness" and "Relationship with neighbors" questions. This shows that the people clustered into "socially connected group" were more inclined and dependent on the connections with other people and external parties and therefore, they had increased their interactions with others even during a pandemic situation. In other words, people in "socially connected group" have strengthened and increased their social connections over the pandemic time in a considerably different manner compared to the people in the "socially reserved group." This is further established considering that the questions in which these considerable differences are present also happen to be the two questions grouped under the "Peer Interactions / Connections" factor.

Furthermore, we can observe a polar opposite behavior within "socially connected group" considering the comparison in Figure 10 and in the mean answers given by the overall population (the results of the entire processed dataset) as shown in Figure 4. Even if the mean behavior of the entire population is closely matched with the responses of the "socially reserved group," in comparison, the direction of the mean values which the "socially connected group" had provided shows a polar opposite behavior. As an example, the overall respondent population showed a general decrease in responses related to "Social connectedness" and "Relationship with neighbors" whereas upon clustering, it was identified that the respondents clustered under "socially connected group" showed an increase. Therefore, we can further conclude that the unsupervised clustering has resulted in "more connected" and "more reserved" people being classified under the two different groups.

Even when considering the differences in values shown in Figure 10, corresponding to the use of social media and hobbies as well as feelings such as loneliness, emptiness and fear of death, it is clear that the individuals clustered under "socially connected group" have experienced a lesser need to depend on coping mechanisms during the pandemic. In addition they have also indicated a comparative reduction in the heightened symptoms related to emotional distress owing to the lockdown. This could be a result of them being more socially connected in comparison to the other cluster while also considering social connectedness as a method of dealing with the changes in lifestyles that came about with the lockdowns and travel restrictions (Bellis et al., 2020; Goodwin et al., 2020). In comparison, the more "socially reserved group" has shown a tendency toward the consumption of social media and at the same time their feelings such as loneliness, emptiness and fear of death have increased compared to the more "socially connected group." It could be speculated that the response of the more "socially connected group" to the drastic changes that were brought about as a result of the pandemic relied on making efforts to maintain connections with others (Cooper et al., 2021; Long et al., 2022). This behavior of data further proves that "socially connected group" sets itself apart from the majority of the population in terms of its psychological response to the pandemic, showcasing a clearly identifiable contrast in their dependency on social interactions to maintain their mental well-being. Considering all the above results and the differences in responses, it is evident that the individuals belonging to "socially connected group" have experienced an increase in maintaining "Peer Interactions / Connections" (factor 4 in Table 4) in comparison to "socially reserved group" even during the pandemic period. This supports the naming of Group 01 as "socially connected group" and Group 2 as "socially reserved group" in relation to their dependency and ability on interactions with the surrounding society. Our study's findings are consistent with previous research, which shows that social connections, and being more "Connected" in nature contribute to better psychological well-being (Mehl et al., 2010; Okabe-Miyamoto et al., 2021).

To summarize, this research study aimed at finding the psychological impact of the COVID-19 pandemic in the Sri Lankan context using a data-driven approach. Along with the confinement to homes and isolation from usual social interactions, the impact on the mental health of individuals was inevitable. The closure of schools and workplaces for prolonged periods, often indefinitely, resulted in making a major switch to online platforms where almost all interactions and activities had to be carried out remotely. As a result of this increased dependency on devices and constant exposure to virtual environments, people belonging to almost all demographics in Sri Lanka experienced emotional distress and negative psychological repercussions. Feelings such as loneliness and emptiness were heightened and addiction to social media and substance abuse were also observed during the pandemic.

Due to its heterogeneous population and range of resource-access constraints affecting the population, Sri Lanka is a befitting place to investigate the effects of COVID-19 on critical dimensions, including psychological effects, as influenced by socioeconomic variables. Sri Lanka and other low-income nations have been particularly hard hit by pandemic's psychological effects. With the complete dependency on social media and other forms of digital content for updates regarding the pandemic and access to health information, people often fall victim to fake news being circulated and the constant reminders and focus given to daily death counts and infection statistics through media, contributed considerably to causing emotional distress and feelings such as anxiety and fear among people (Adnan et al., 2022). Receiving word of relatives and close contacts being infected with the virus also factored into these negative repercussions to mental well-being. The minimal human interactions that were permitted for individuals being subjected to quarantine, whether at home or at quarantine centers, were also factors that affected their psychological state (Buecker and Horstmann, 2022). Being deprived of usual human interactions and mobility being limited to the bare essentials were drastic changes that occurred in lifestyles around the world, which was challenging in terms of maintaining relationships crucial to the mental well-being of human beings, who are social creatures by nature (Scala et al., 2020).

Specifically considering the context of the pandemic period where lockdowns were imposed and people were often confined to their homes, it is understandable to assume that the mental state of individuals heavily depended on their ability to maintain interpersonal relationships. As made clear by the findings of the analysis performed, a certain group of individuals were more comfortable with the concept of self-isolation and reduced number of connections to the external society. Their satisfaction could be sufficiently derived from minimal interactions limited to immediate family members and other close contacts if necessities such as food and shelter were provided as per their needs.

However, in contrast, there also exists a distinct cluster of individuals ("socially connected group") who seek solace and comfort in interacting with peers that often are found beyond the circle of people which includes family members and close friends. Therefore, the members of the "socially connected group" of people relied a lot less on social media and hobbies and reached out to others lowering their fear of death along with loneliness and emptiness levels and this is also in line with recent studies conducted regarding the psychological effects of the pandemic (Pedrosa et al., 2020; La Rosa et al., 2024). They were often inclined to seek quality in the relationships maintained, which translates to preferring real-time interactions over social media where people were asynchronously divided by screens, which could have served as a method for them to respond to their psychological needs in keeping a healthy state of mind amidst the chaos that surrounded the virus spread and the detachment from usual routines.

This contrast in preferences could thus be circled back to the previous analysis on the "connected" and "reserved" nature of humans. The variations in perception of social connectedness, whether it be relationships with family, neighbors or office colleagues in addition to feelings displaying emotional distress symptoms, were unique identifiers which largely depended on the individual respondents. This could be presented as a key finding of the research study important in understanding the nature of human behavior and their psychological responses to the pandemic. As mentioned at the end of the Results section, it was also noteworthy that there was no demographic marker that distinguished between these two groups such as income level, education level, residing state, ethnicity, age, gender, marital status or education level and the only clear differentiator between the two groups that resulted in unsupervised clustering was whether the population groups were more socially connected or not.

## 5 Limitations

The nationwide door-to-door, face-to-face, CAPI (Computer-Assisted Personal Interview) field survey (Ilangarathna et al., 2023) conducted covering 3,020 households in Sri Lanka, which the data for this study was collected, was conducted focusing on retrieving information on seven different areas (refer to the previous sections) about how the population was impacted during the COVID-19 pandemic. Therefore, the entire study spanned questions not just covering the psychological impact, but also on other six focus areas. As such, questions were balanced between all impact areas and households from all nine provinces in Sri Lanka were selected for this survey as mentioned in the previous sections. We recognize that another similar survey could be conducted with more questions to capture the psychological impact alone during the COVID-19 pandemic and such an expansive survey might present results with greater resolution. However, such a survey will cost even more than the survey which was conducted by us. If more similar surveys had been done with their survey responses made publicly available, a better comparison between Sri Lanka and other countries could have been conducted. Furthermore, this study was conducted in a data-driven approach and not based on a prior hypothesis to not be biased by any judgment on how the population might have behaved during the pandemic. Using this approach, we were able to gain insights purely based on factor analysis and unsupervised clustering. However, authors acknowledge that the questionnaire design may carry inherent bias and limitations due to the selection of questions. While factor analysis was employed to minimize redundancies and misalignments quantitatively, some residual bias may still persist, reflecting the inherent constraints of such methodologies. We have presented the data analysis techniques used in this study above in the previous sections and since the data used in this study are publicly available, we invite the research community to further explore and draw insights from the data using other techniques as appropriate.

## 6 Conclusion

The study reveals that the psychological impact of the COVID-19 pandemic in Sri Lanka was heavily influenced by the nature of social connectedness and coping mechanisms adopted during periods of lockdown and isolation. Through factor analysis and unsupervised clustering, two distinct groups were identified. Examining the responses of the individuals categorized into these two groups, those who relied on interpersonal relationships to maintain mental well-being were termed as "socially connected group" and those who were more self-sufficient and preferred minimal interactions were termed "socially reserved group." The "socially reserved group" was more inclined to use social media and hobbies as coping mechanisms as opposed to reaching out to make connections with friends, family and neighbors. Hence, they had higher levels of loneliness and emptiness compared to the more "socially connected group" of people. Another key observation of the factor analysis is that the population did not view social media as a means of peer interaction or maintaining human connections, but rather as a coping mechanism.

As demonstrated in this study, in future scenarios involving prolonged isolation or similar pandemic circumstances, the mental well-being of individuals may be significantly influenced by their social tendencies depending on whether they are more reserved or connected. Another interesting conclusion that can be drawn from the unsupervised clustering was that the two groups were not distinguishable depending on their demographics such as income level, education level, residing state, ethnicity, age, gender, marital status or education level. It would be beneficial to raise awareness regarding this in order to make sure that the community is well-equipped to provide necessary assistance when people are in need during a similar pandemic situation. For example, organizing socializing events via online platforms or at a physical setting (adhering to imposed health regulations) could be helpful in maintaining the psychological well-being of individuals by ensuring that each person receives their share of quality peer interactions.

Using these results stakeholders responsible for mental health, including but not limited to government agencies, policymakers, and counseling practitioners can draw valuable insights to better tackle issues that may arise in the future. This research can serve as a guide to inform and refine their strategies, ensuring more targeted and effective mental health support during such challenging times. Additionally, this work offers guidance for future, in-depth research attracting additional validation from larger academic and policy-making communities, particularly also in the formulation of policies relevant to the use of social media.

## Data availability statement

The original contributions presented in the study are included in the article/supplementary material, further inquiries can be directed to the corresponding author/s.

## Ethics statement

The studies involving humans were approved by Ethical Review Committee, Faculty of Arts, University of Peradeniya. The studies were conducted in accordance with the local legislation and institutional requirements. Written informed consent for participation in this study was provided by the participants' legal guardians/next of kin.

## Author contributions

IT: Conceptualization, Data curation, Formal analysis, Investigation, Methodology, Validation, Visualization, Writing – original draft, Writing – review & editing. TF: Conceptualization, Data curation, Formal analysis, Methodology, Validation, Visualization, Writing – original draft, Writing – review & editing. IM: Conceptualization, Data curation, Methodology, Validation, Writing – original draft, Writing – review & editing. RG: Conceptualization, Data curation, Investigation, Methodology, Supervision, Validation, Writing – review & editing. VH: Investigation, Supervision, Writing – review & editing, Conceptualization, Methodology. RT: Supervision, Writing – review & editing. AR: Supervision, Writing – review & editing. PE: Conceptualization, Investigation, Methodology, Supervision, Writing – review & editing. JE: Supervision, Writing – review & editing, Investigation, Project administration.

## Funding

The author(s) declare that no financial support was received for the research, authorship, and/or publication of this article.

## Conflict of interest

The authors declare that the research was conducted in the absence of any commercial or financial relationships that could be construed as a potential conflict of interest.

## Publisher's note